\documentstyle[epsfig,floats,amssymb,aps,prl,twocolumn]{revtex}

\begin{document}
\title{Formation of Ramsey Fringes in Double Bose-Einstein Condensates}
\author{A. Eschmann, R.J. Ballagh and B.M. Caradoc-Davies,}
\address{Physics Department, University of Otago, Dunedin, New Zealand.}
\wideabs{
\maketitle
\vspace{1cm}
\begin{abstract}
Numerical simulations of Ramsey fringe formation in double Bose condensates
are carried out in two spatial dimensions. The effects of  meanfield
nonlinearity and diffusion  in the condensates give rise to new features in
the Ramsey fringes, namely spatial variation  and an asymmetry between
positive and negative field detunings. We introduce the concept of an
effective detuning, which for moderate interpulse times provides a
qualitative understanding of the new features.
\end{abstract}

}

\section{Introduction}

Recent experiments carried out by the JILA and MIT groups$^{\cite{jila},\cite
{jilaexp}}$ have shown the rich behaviour that can occur in double
condensate systems. In particular the JILA\ group have used Ramsey's$^{\cite
{ramsey}}$ separated pulse technique to produce interference fringes which
demonstrate the long term phase coherence of their condensates. Although
Ramsey's original theoretical description explains qualitatively the
behaviour that is seen, additional mechanisms present in condensates, namely
nonlinearity from the mean field collisions and diffusion due to kinetic
energy, complicate the condensate case. In this paper, we present a
theoretical study of Ramsey fringe formation in two spatial dimensions,
including the effect of the mean field collisions and kinetic energy. The
JILA experiment minimises the nonlinear  effects by only partially
overlapping the condensates, so that the experiment is carried out in the
low density (linear) region. Here, we consider the case where the
condensates are fully overlapped, and show that the additional condensate
mechanisms can give rise to spatial variation of the fringes across the
condensate, and  to an asymmetry between positive and negative detunings. 
Eventually, for large separation times between pulses, the regularity of
fringes predicted by the original theory is disrupted by the additional
mechanisms.

\section{Formulation}

In Ramsey's original work$^{\cite{ramsey}}$ a beam of atoms in the ground
state was passed through two separated oscillatory field regions. The
interaction with the field in the first region transfers some population
into the excited state and creates a magnetic dipole. Between the
oscillating fields the atomic populations remain fixed, but the dipole
undergoes free precession. The interaction with the second field causes an
additional transfer of population, which is sensitively dependent on the
total phase accumulated by the dipole in the field free region. If the
pulses are separated by a fixed time $T$, the measured population shows a
fringe pattern (as a function of the oscillating field frequency) known as a
Ramsey fringe, which is analogous to that obtained in a 2-slit interference
experiment. The experiment can also be carried out with a fixed field
frequency and varying pulse separation times $T,$ which leads to a similar
interference pattern, with the population varying according to $\cos (\Delta
T)$, where $\Delta $ is the atom-field detuning.

In the double BEC system, the two atomic states of the original Ramsey
system become the two condensate states of different internal states, and
the oscillating fields are 2-photon $\pi /2$ pulses (of duration $\tau =\pi
/(2\Omega )$, where $\Omega =\sqrt{\Omega _{R}^{2}+\Delta ^{2}}$, and $%
\Omega _{R}$ is the 2-photon Rabi frequency$^{\cite{dum}}$). We choose in
this paper to scan the pulse separation time $T$ and hold the detuning
constant, following the procedure used in the JILA experiment$^{\cite
{jilaexp}}$

In the meanfield limit, the double condensate system can be described by the
coupled Gross-Pitaevskii (GP) equations$^{\cite{lifschitz},\cite{ballagh}}$,

\begin{eqnarray}
\frac{\partial \psi _{1}({\bf r},t)}{\partial t} &=&i\nabla ^{2}\psi _{1}(%
{\bf r},t)-{\frac{{ir^{2}}}{{4}}}\psi _{1}({\bf r},t)+iV\psi _{2}({\bf r},t)
\nonumber \\
&&-iC[|\psi _{1}({\bf r},t)|^{2}+w|\psi _{2}({\bf r},t)|^{2}]\psi _{1}
({\bf r},t)  \label{GPE1} \\
\frac{\partial \psi _{2}({\bf r},t)}{\partial t} &=&i\nabla ^{2}\psi _{2}(%
{\bf r},t)-{\frac{{ikr^{2}}}{{4}}}\psi _{2}({\bf r},t)+iV\psi _{1}({\bf r}%
,t) \nonumber  \\ &&-i\Delta \psi _{2}({\bf r},t) \nonumber \\
&&-iC[|\psi _{2}({\bf r},t)|^{2}+w|\psi _{1}({\bf r},t)|^{2}]\psi _{2}
({\bf r},t).  \label{GPE2}
\end{eqnarray}
where the spatial and temporal coordinates, ${\bf r}$ and $t$ , are scaled
as in Ballagh et al $^{\cite{ballagh}}$. The nonlinearity parameter $C$ is
proportional to the total number of atoms $N$ in the condensate and the
intraspecies {\em s}-wave scattering length, while the factor $k$ indicates
the relative trapping strength of  state $|2\rangle.$ In this paper,
 we take $k=1$,
  so that both components are equally trapped, as in the JILA experiment.
The factor $w$ gives the ratio of inter- to intra- species scattering
length, and includes the effect of wavefunction symmetry. We have assumed
that the intra-species {\em s}-wave scattering length $a_{ii}$ is the same
for each internal state of the atom, so that $a_{11}=a_{22}.$ This is
reasonable for the alkali atoms, and makes the $C$ values the same in both (%
\ref{GPE1}) and (\ref{GPE2}), but our results could easily be generalised to
situations of unequal intra-species scattering length. Growth and loss
processes are ignored here, so that $N$ is conserved. The coherent coupling
between condensate components is described by the terms containing $V$ and $%
\Delta $ , where  $V$ is half the bare Rabi frequency, and $\Delta $ is the
detuning of the coupling field from the bare atomic
resonance$^{\cite{detune}}$.

It is useful at this point to introduce a new quantity $\Delta _{eff}$,
which is the difference between the diagonal terms in Eqns (\ref{GPE1}) and (%
\ref{GPE2}), 
\begin{equation}
\Delta _{eff}({\bf r,}t)=C(w-1)(|\psi _{1}({\bf r},t)|^{2}-|\psi _{2}({\bf r}%
,t)|^{2})+\Delta .  \label{eff_det}
\end{equation}
This quantity, which we will interpret as an effective detuning, will prove
helpful in interpreting the effect of the nonlinear mechanism on fringe
formation. It is clear that when $w=1$ (all scattering lengths the same), or 
$|\psi _{1}({\bf r},t)|^{2}\approx |\psi _{2}({\bf r},t)|^{2}$ (near the
edges of the wavefunctions), $\Delta _{eff}({\bf r},t)$ reduces to the usual
value of the detuning $\Delta $.

\section{Results}

Equations (\ref{GPE1}) and (\ref{GPE2}) are solved for a succession of $T$
values at fixed detuning $\Delta $, using a modified split-step Fast Fourier
Transform method. Initially all the population is in component 1, which we
choose to be an eigenstate of the uncoupled single component system. A grid
of 512x512 points is used for the simulations, over a spatial range of $%
-20\leq x,y\leq 20$. In our results, we have chosen the case of $w=2,$ which
makes apparent the role of the meanfield nonlinearity, and is realistic for
sodium.

\begin{figure}
\begin{center}
\epsfbox{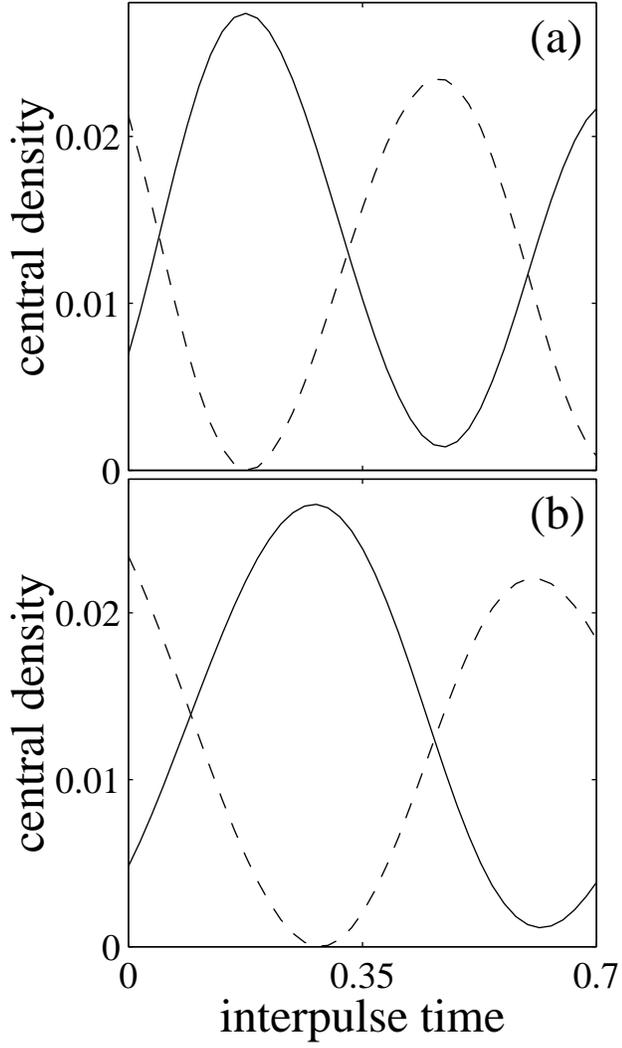}
\end{center}
\caption{Condensate density of component 1 (solid line) and component 2
(dashed line) at the trap centre, as a function of interpulse times $T$. (a)
$\Delta =10$ (b) $\Delta =-10$. The density is taken immediately following
the second pulse. Parameters are $V=10$, $C=200$, $w=2$, $\tau=0.0703$} 
\label{centrefringe}
\end{figure}

\begin{figure}
\begin{center}
\epsfbox{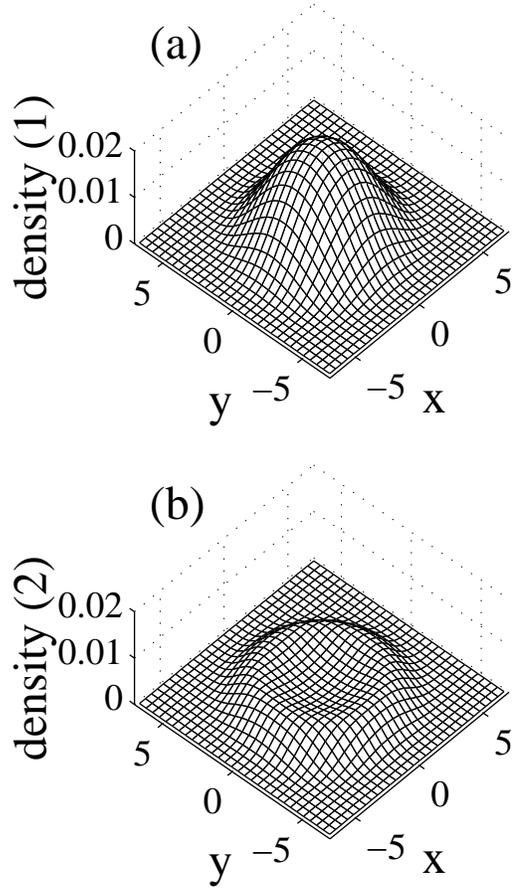}
\end{center}
\caption{Spatial distribution of condensate density of (a) component 1
 (b) component
2,  for an interpulse time of
$T=0.6674$. Parameters are as in Fig.~\ref{centrefringe}(a) }
\label{surface}
\end{figure}

\begin{figure}
\begin{center}
\epsfbox{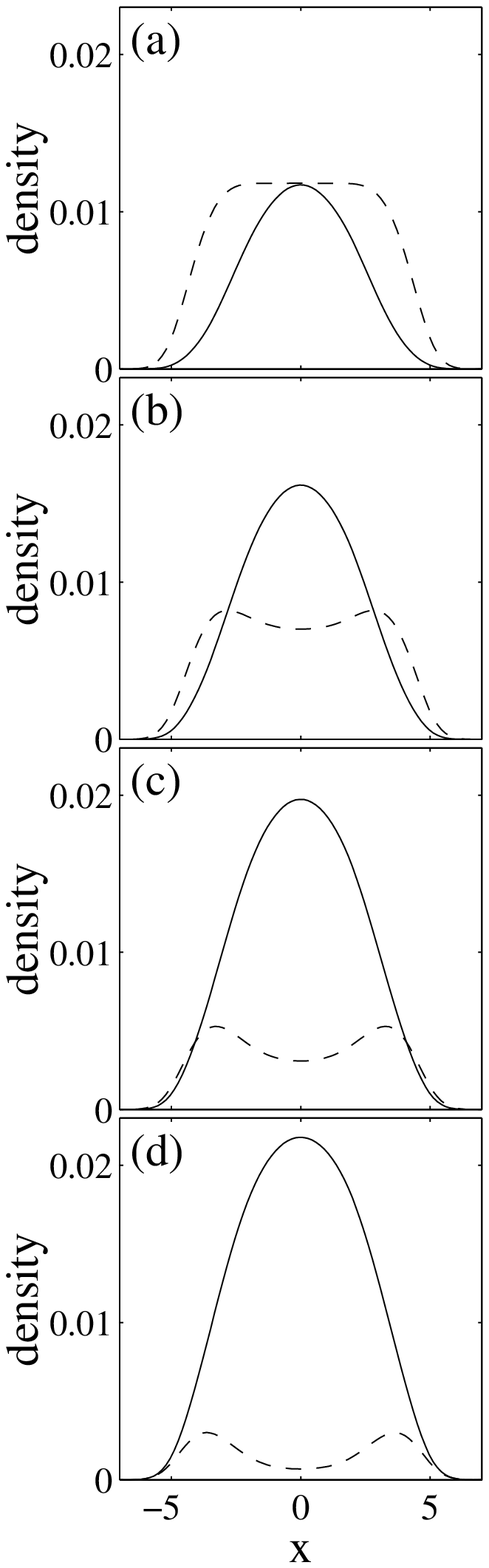}
\end{center}
\caption{ Condensate density of component 1 (solid line) and component
2 (dashed line) along the $x$ axis, for  a sequence of interpulse times (a)
$T=0.5971$ (b) $T=0.6322$ (c) $T=0.6674$ (d) $T=0.7025$. The detuning is 
$\Delta =10,$ and all other parameters are as in Fig.~\ref{centrefringe}.
Note that (c) corresponds to Fig.~\ref{surface}.} 
\label{pos_slice}
\end{figure}

\begin{figure}
\begin{center}
\epsfbox{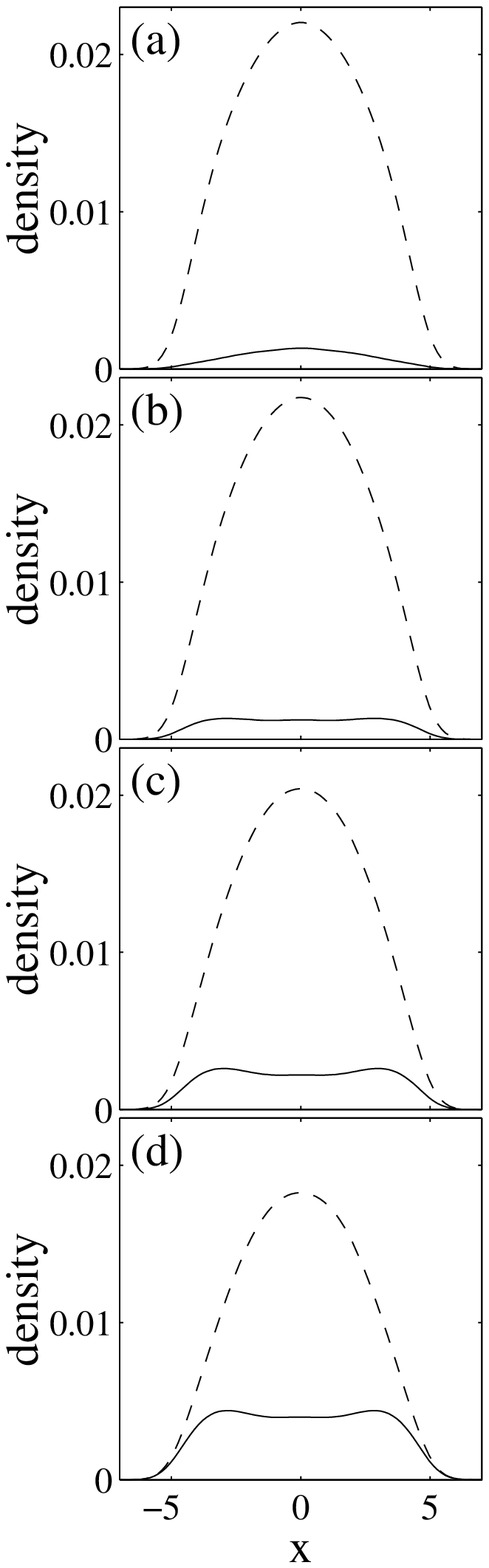}
\end{center}
\caption{Condensate density of component 1 (solid line) and component
2 (dashed line) along the $x$ axis,  for a sequence of interpulse times (a) 
$T=0.5971$ (b) $T=0.6322$ (c) $T=0.6674$ (d) $T=0.7025$. The detuning is 
$\Delta =-10,$ and all other parameters are as in  Fig.~\ref{centrefringe}.} 
\label{neg_slice}
\end{figure}

A typical result is shown in Fig.~\ref{centrefringe}(a), where the central 
condensate density (i.e. at $x=0,$ $y=0$ ) of
each component immediately following the interaction with the second
 oscillatory field is plotted as a function of the
interpulse time $T.$ The figure displays the characteristic Ramsey
fringe pattern, however more careful inspection of our results shows that
the additional condensate mechanisms cause significant modifications to the
fringe behaviour. This is most obviously apparent in the
spatial density  distributions of each component,
and in Fig.~\ref{surface} we plot the full two dimensional behaviour 
for the same case as Fig.~\ref{centrefringe}(a), at the interpulse time of 
$T=0.6674$. Spatial structure is now very evident in 
component 2, and this 
can not be accounted for in the simple Ramsey theory. Density slices
taken across the $x$ axis allow a more quantitative evaluation 
of the spatial structure, and in Fig.~\ref{pos_slice} we plot, 
for the same  simulation parameters 
as in Fig.~\ref{centrefringe}(a), 
a sequence of density slices corresponding to a sequence of
equally spaced interpulse times. 
It is clear from this sequence that component 2 does
not replicate the initial eigenstate of component 1, but instead develops a
dip in its centre, which increases as $T$ increases.
A further significant difference of the condensate case
is seen when we consider the effect of changing the sign of the detuning $%
\Delta .$ Ramsey's original theoretical expression for the fringes is
exactly symmetric to the change $\Delta $ $\rightarrow -\Delta .$ In the
condensate case,  when we repeat the previous calculations with the opposite
sign for detuning (and all other parameters unchanged), we obtain
significantly different results, as shown in Fig.~\ref{centrefringe}(b),
and even more dramatically in Fig.~\ref{neg_slice}.

Comparing Figs.~\ref{centrefringe}(a) and (b). we see that the period of
the Ramsey fringe at the centre of the wavefunction is distinctly different
for detuning of opposite signs: in Fig.~\ref{centrefringe}(a)
 (where $\Delta $ $=10$) the
first minimum of component 1 occurs for $T=0.47,$ while in 
Fig.~\ref{centrefringe}(b)) (where 
\mbox{$\Delta =-10$}) the first minimum of component 1 occurs for $T=0.61.$
Comparing Figs. \ref{pos_slice} and \ref{neg_slice}, for which
the sequence of times are identical and
only the sign of the detuning has changed, the contrast is immediately
apparent. In Fig.~\ref{pos_slice}, component 2, which is decreasing as 
a whole during the
sequence, develops a central dip. In Fig.~\ref{neg_slice} on the other
hand, component 1
is increasing as a whole, and develops a central dip.

These results can be understood qualitatively in terms of the effective
detuning $\Delta _{eff}({\bf r},t)$, and the well known Thomas Fermi
approximation. The effective detuning $\Delta _{eff}({\bf r},t)$ is a
dynamic quantity that depends on the instantaneous population density
difference $|\psi _{1}({\bf r},t)|^{2}-|\psi _{2}({\bf r},t)|^{2}.$ Within
the Thomas Fermi approximation (which neglects the $\nabla ^{2}$ terms in
the GP equation) no mixing occurs between different spatial points, and
calculations are greatly simplified. For example the atomic dipole created
by the first pulse can be calculated, with the full dynamic behaviour of $%
\Delta _{eff}({\bf r},t)$ taken into account, by solving an ordinary
differential equation. However, good estimates of the behaviour of the
dipole can be obtained even more simply, as we will now discuss. We note
first that during the interval between pulses, the populations at position $%
{\bf r}$ do not change (as a result of neglecting the $\nabla ^{2}$ term),
and the phase accumulated by the atomic dipole is determined by the value of 
$\Delta _{eff}({\bf r},\tau )$ at the end of the first pulse (which is of 
length $\tau$). The value $%
\Delta _{eff}({\bf r},\tau )$ is determined by the inversion density at time 
$\tau ,$ and for a constant detuning $\Delta $ the inversion following the
first pulse is given by the well known  formula$^{\cite{alleneberly}}$, 
\begin{equation}
w=w_{0}\frac{\Delta ^{2}+4V^{2}\cos \Omega _{0}\tau }{\Omega _{0}^{2}}
\label{inversion}
\end{equation}
where $w_{0}$ is the initial inversion and $\Omega _{0}=\sqrt{4V^{2}+\Delta
^{2}}.$ At the edges of the condensate, where the densities are low, $\Delta
_{eff}$ is essentially $\Delta ,$ and thus the first $\pi /2$ pulse ($\Omega
_{0}\tau =\pi /2)$ produces a resulting inversion of $w_{0}(\Delta /\Omega
_{0})^{2}.$ Elsewhere across the wavefunction, we can approximate the effect
of the dynamically changing detuning by using a temporal average of $\Delta
_{eff}({\bf r},t)$ during the pulse, which we will write as $\bar{\Delta}%
_{eff}({\bf r}).$ At the centre of the condensate, the densities are
largest, and the average over time of $|\psi _{1}({\bf r},t)|^{2}-|\psi _{2}(%
{\bf r},t)|^{2}$ is positive during the pulse, so that $|\bar{\Delta}_{eff}(%
{\bf r})|$ exceeds $|\Delta |$ $($for positive $\Delta$). Thus from Eq.(%
\ref{inversion}) we see (in the regime where $V \lesssim \Delta$) that the
change
in the initial inversion density is  less  at the centre than
at the edges. 
Thus for positive detuning,  $\Delta _{eff}({\bf r},\tau )$ is {\em greater}
at the centre of the wavefunction than at the edges, and thus the subsequent
phase  accumulation between pulses is greater in the centre. We thus expect
that the  Ramsey fringes will have a greater frequency in the centre of the
wavefunction, which provides the explanation for  the behaviour
in Fig.~\ref{pos_slice}.
There we see that  population transfer out of  component 2 is more rapid at
the centre than at the edges, which means the interpulse minimum is reached
first at the centre.  In the case of negative $\Delta $, the magnitude of
the effective detuning will be {\em less }at the centre than at the edges
[provided $|\Delta |$ is larger than $|C(w-1)(|\psi _{1}({\bf r}%
,t)|^{2}-|\psi _{2}({\bf r},t)|^{2})|$]. This means the frequency of the
Ramsey fringes will be less at the centre of the wavefunction than the
edges, and explains why, in Fig.~\ref{neg_slice}, the edges 
of component 1 reach their
maximum more rapidly than the centre. It is also now clear that Ramsey
fringes at the centre of the wavefunction will have a greater frequency for
positive detuning than for negative detuning, which indeed is the behaviour
that is revealed by comparison of  Figs.~\ref{centrefringe}(a) and (b).

\begin{figure}
\begin{center}
\epsfbox{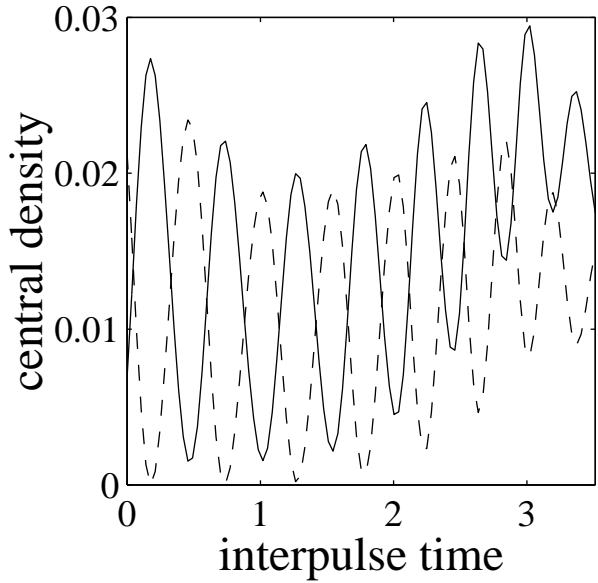}
\end{center}
\caption{Condensate density of component 1 (solid line) and component
2 (dashed line) at the trap centre, as a function of interpulse times $T.$
All  parameters are as in Fig.~\ref{centrefringe}(a)} 
\label{random}
\end{figure}

Finally we present, in Fig.~\ref{random}, the behaviour of the
 Ramsey fringes for a large
range of interpulse times $T.$At long times, the effects of diffusion
combined with meanfield nonlinearity  produce irregular periodicity. We note
too that a plot of the spatial dependence of the population densities at a
given large value of $T$ will usually show significant modulation. Once
diffusive mixing becomes significant, the predictions of our effective
detuning, which relies on the Thomas Fermi approximation, are no longer valid.
The fringe behaviour must then be obtained from a full solution of the coupled
GP equations (\ref{GPE1}) and (\ref{GPE2}).

\section{Conclusions}

We have carried out numerical simulations in two spatial dimensions of
Ramsey fringe formation in double Bose condensates. Our calculations include
the effects of both nonlinearity and diffusion, and in contrast to the
reported JILA\ experiment, we have considered the case of fully overlapping
condensates, which enhances the effect of nonlinearity. The mechanisms of
diffusion and nonlinearity give rise to new features in the Ramsey fringes,
namely spatial variation of the fringes, and an asymmetry between positive
and negative field detunings. By introducing the concept of an effective
detuning we have obtained,  for moderate interpulse times, a qualitative
understanding of the new features. However the effective detuning accounts
only for  nonlinear effects and eventually, at long enough interpulse times,
diffusion  becomes important and  the full GP equations must then be solved.

\subsection*{Acknowledgments}

This work was supported by the Marsden Fund under contract PVT-603, and by
FRST under contract U00613.


\begin{references}
\bibitem{detune} For two-photon coupling $\Delta$ is the two-photon detuning.
\end{references}
\end{document}